# Certain Interesting Properties of Action and Its Application Towards Achieving Greater Organization in Complex Systems

Atanu Bikash Chatterjee

Department of Mechanical Engineering, Bhilai Institute of Technology, Bhilai House
Durg, 490006 Chhattisgarh, India
e -mail: abc3.14160@yahoo.com

**Abstract:** The Principle of Least Action has evolved and established itself as the most basic law of physics. This allows us to see how this fundamental law of nature determines the development of the system towards states with less action, *i.e.,* organized states. A system undergoing a natural process is formulated as a game that tends to organize the system in the least possible time. Also, other concepts of game theory are related to their profound physical counterparts. Although no fundamentally new findings are provided, it is quite interesting to see certain important properties of a complex system and their far-reaching consequences.

**Keywords:** Action, dead state, extensive property, super-structure, system, organization, complexity, Nash-equilibrium strategy, pay-off function.

**Introduction:** All processes occurring in the universe are rooted in physics and have a physical explanation. All the structures in the universe exist, because they are in their state of least action or tend towards it. All natural process are directed along the steepest descents of an energy landscape by equalizing differences in energy *via* various transport and transformation processes, *e.g.,* diffusion, heat flows, electric currents and chemical reactions [1]. In any system, simple or complex, the system spontaneously calculates which path will use least effort for that process. In an organized system, the action of a single element will not be at minimum, because of the constraints, but the sum of all actions in the whole system will be at minimum [2]. The action of a single element is not maximal as well, because by definition this will destroy the system, so this intermediate state represents an optimum. The reorganization of the system to achieve it is a process of optimization. A system comprises of elements and constraints, both internal as well as external. The internal constraints could be the configurations of the system or the elements themselves, whereas, the external constraints define the geometry of the system. The elements apply work on the constraints to modify the organization and minimize the action, which takes finite amount of time, making reorganization a process. The state of the constraints in the system that determines the sum of the actions of its elements is called organization. The dynamical systems that are present in nature are generally very complex with various levels of complexity present within themselves. These systems show the property of emergence. Generally, those systems that are observed for a long span of time show emergent simplicity as they are continuously reorganizing themselves towards a state of greater order. Order implies a state of least action. An interesting property of a complex system as quoted by Georgi Georgiev in his paper states, "the minimum of the sum of all actions is not necessarily the minimum of the action of each element, but some optimal value" [2]. This leaves us with an important question; can this optimal value of action of a system ever assume an infinite, negative or null value? A complex system with a structure and emergence is said to be self-organizing. So the process of self-organization of the systems can be called a "Process of achieving a least action state by a system" [2]. It could last billions of years or indefinitely. When John von Neumann formulated game theory [3] as a mathematical model of human behaviour, he drew inspiration from the behaviour of thermodynamic systems. Also, when John F. Nash [4, 5] expanded the theory with a solution concept for two or more players, he had in mind the chemical equilibrium of many substances. Today, applications of game theory extend from descriptions of biochemical and biophysical processes to accounts of impressive breadth of phenomena, most notably in economics, biology and social sciences as well as in engineering and computer and information sciences. Therefore, could it be, just as the two pioneers envisioned, that there is, after all, a profound correspondence and not only a mere resemblance between human behaviour and physical processes? [6] Another intriguing question is that whether the games played by the particles within the system or the system itself with its surrounding, are cooperative or non-cooperative? And does there exist any optimal strategy or a Nash equilibrium strategy in achieving the states of least action? This paper and other papers in future are going to provide a better understanding and focus on these and more similar questions.

**Methodology:** A process can be defined as an act by which a system tends to stabilize or organize itself with passage of time. Any process occurs in nature due to the existence of a positive gradient of the energy between the system and its surrounding universe. This gradient is measured in terms of the quality of the energy the system possesses hence, its <u>exergy</u> [7, 8, 9]. This energy gradient acts as a driving force enabling a dynamical system to organize itself with continuous evolution of time.

The principle of least action states that the actual motion of a conservative dynamical system between two points occurs in such a manner, that the action has a minimum value in respect to all other paths between the points, which correspond to the same energy.

The classical definition of the principle of least action [10, 11, 12, 13] is:

$$\Delta \int_{t_1}^{t_2} p_i q_i \, dt = 0$$

The variation of the path is zero for any natural process occurring between two points of time $t_1$ and $t_2$, or nature acts in the simplest way hence, in the shortest possible time.

The action:

$$I = \int_{t_1}^{t_2} L \, dt$$

Where L is the lagrangian of the system:
L=T-V

Here, T and V are the kinetic and the potential energy of the system respectively. For the motion of the system between time $t_1$ and $t_2$, the lagrangian, L, has a stationary value for the correct path of motion. This can be summarized as the Hamilton's Principle [13].

$$\delta I = \delta \int_{t_1}^{t_2} L \, dt$$

For an N-element system, this can be generalized as [2]:

$$\delta \left( \sum I_i \right) = \delta \left( \sum \int_{t_1}^{t_2} L_i \, dt \right)$$

Where, $\sum I_i = I_1 + I_2 + I_3 + I_4 + \ldots\ldots\ldots\ldots\ldots I_N$
$I_1, I_2, I_3\ldots$ represent the action of individual elements within a system.

**Case-1:**
When the sum of the action of the system or action of any individual element assumes a negative value, *i.e.,* $(\sum I_i) < 0$ or $I_i < 0$.
The system or that particular element tends to disorganize itself with passage of time. Re-organization with passage of time can be thought of as a direction of flow of time. Time cannot flow backwards. Re-organization tends to reduce the potential of a system compared to its surroundings hence, degrading its exergy. Disorganization is against the law of nature causing the system to collapse immediately. Thus, it is impossible for a system or any system element to possess negative value of action.

**Case-2:**
When the action of the system or the action of any individual element of the system approaches infinity, *i.e.,* $(\sum I_i) \to \infty$ or $I_i \to \infty$.
An organized system tends to have the least value of action. Conversely, lesser is the value of action more is the amount of organization present in a system. Thus, amount of organization (Org) is inversely related to the action of the system (I).

$$(\text{Org}) \times (I) = \text{constant}$$

Differentiating with respect to time,

$$\frac{d\text{Org}}{dt} = -\frac{\text{Org}}{I} \times \frac{dI}{dt}$$

[3]

The equation clearly shows that, the rate of increase of organization in a system is inversely related to rate of decrease in action of the system multiplied by the ratio of amount of organization to the amount of action present in the system.

When $I \to \infty$,

$$\frac{dOrg}{dt} = 0$$

This implies a state of the system, where amount of organization does not depend on the time variable any more. But at this point of time, organization also approaches zero value as action and organization are related inversely. Thus, such a system, again fails to exist in the universe as the system would be highly disorganized.

**Case-3:**

If the sum of the action of a system becomes zero, *i.e.*, $(\sum I_i) = 0$.

This case gives rise to an important property of action.

Summation of action of all elements is represented as:

$$\sum I_i = I_1 + I_2 + I_3 + I_4 + \ldots\ldots\ldots\ldots\ldots I_N$$

$$(\sum I_i) = 0$$

So, $\quad I_1 + I_2 + I_3 + I_4 + \ldots\ldots I_N = 0$

Where,

$$I_N = \int_{t_1}^{t_2} L_N dt$$

From the above cases it is clearly seen that, for a system to exist in the universe and undergo a process of self-organization, it is necessary that the action assumes a positive non-infinite value.

$$(\sum I_i) = (\sum \int_{t_1}^{t_2} L_i dt) = 0$$

Where, the system undergoes a process between times $t_1$ and $t_2$.

[4]

From the property of definite integral,

$$(\sum I_i) = (\sum \int_{t_1}^{t_2} L_i dt) = (\sum (\int_{t_1}^{\alpha} L_i dt + \int_{\alpha}^{t_2} L_i dt)) = 0$$

Where, $\alpha$ is an intermediate time between $t_1$ and $t_2$. From the above expression it can be clearly inferred that the individual integrals assume zero value thus, giving us a root between $t_1$ and $t_2$, the root being "$\alpha$".
So, $L(\alpha) = 0$.

This analysis can be again done taking another set of values between $(t_1, \alpha)$ and $(\alpha, t_2)$.

$$(\sum I_i) = (\sum \int_{t_1}^{t_2} L_i dt) = (\sum (\int_{t_1}^{\alpha} L_i dt + \int_{\alpha}^{\mu} L_i dt + \int_{\mu}^{\beta} L_i dt + \int_{\beta}^{t_2} L_i dt)) = 0$$

Where, $t_1 < \alpha < \mu < \beta < t_2$ and the individual set of action being zero.
If such infinite iterations are carried out between $t_1$ and $t_2$, then we are left with infinite roots between the set of time interval. Each point within $(t_1, t_2)$ is a root of the action function. This shows that, when the system shrinks to a singular point, then the total action of the system or any individual element within the system vanishes. The action no longer depends on time and assumes a constant value, *i.e.,* a null value.
This clearly shows that action is a property of an evolving system, and depends on the structure of the system, hence, action must be an extensive property [7, 8].
In the earlier case, it was shown that action varies inversely with the amount of organization present within the system. So, at a state with zero action, amount of organization approaches infinity. Approaching a more organized state is the natural tendency of a system. Independent of the instantaneous configuration, the system continuously reconfigures itself with passage of time. Reconfiguration causes the sum total of the action of the system and the action of the independent elements to change continuously but the motive or the strategy of the system remains unique, *i.e.*, to get more and more organized with time. Organization is, thus clearly an intensive property [7, 8] of the system.
But, when the action becomes zero, as in this case, either the system collapses by shrinking to an infinitesimal point or it attains a state of maximum organization. Attaining a state of maximum organization implies the disappearance of the energy gradient between the system and its surrounding. Such a system is said to have reached a dead state [7, 8, 9] where all natural processes have ceased to exist. A natural question arises here:
What is the fate of a system as it reaches an absolute dead state?
As it was stated earlier, that all natural processes cease to occur at the dead state, the system after reaching the dead state, no longer evolves with time but has become a static structure. The system would continue to remain at that state for an infinite period of time. A least action state is also a state of least amount of free energy [1]. A system's configuration determines the amount of free energy it possesses. A system with zero action must then have no free energy, and hence, no configuration. This implies that after achieving the dead state, the system begins to shrink to a point, or more precisely, both the processes occur almost simultaneously.

In order to analyse the question more deeply, a system with N-elements undergoing a natural process is thought of as a group of N-persons playing a zero-sum non-cooperative game [4, 5]. In a zero sum non-cooperative game, each player's loss is the other player's gain. The players are associated with a set of strategies and corresponding to each strategy, there is a pay-off function. The aim of the players is to maximise their pay-off. Each of the elements has a set of pure strategies denoted by $S_i$ where $S_i = (s_1, s_2, s_3....s_n)$. Corresponding to the set of strategies there is a pay-off function, $\rho_i$, that maps the set of all n-tuples of pure strategies into real numbers. The strategy profile, $S_i$, of an $i^{th}$ element denotes the set of strategies that enables it to attain the state of least action. The pay-off in this case is the amount of organization of this system.

The natural tendency of the system would be to maximize its state of organization, hence, there exists a strategy profile, $S_i^* = (s_1^*, s_2^*, s_3^* ....s_n^*)$, that maximizes the pay-off and hence the organization of the system, represented as,

$$Org = \rho_i (s_1^*, s_2^*, s_3^* ....s_n^*)$$

The strategy profile, $S_i^*$, then represents the Nash-equilibrium [5] for the system.
The Nash-equilibrium strategy profile is unique for a system; it is also an unknown strategy for the system, when it is interacting with its surroundings and also for the elements within the system. It is actually the strategy that the system or the elements of the system are searching for. If this strategy becomes known to the system or to its elements then the system reaches a dead state.

During all the analysis done earlier, an important point was not taken into consideration; In a multi-element system, the elements themselves behave as constraints and thus, block each other's path causing the action of the system to increase and organization to decrease. As a result, Nash-equilibrium is never reached. Furthermore, the shrinking of the system to a point at the state of zero action tends to increase the action to such an amount that the system disintegrates at its various levels of complexity and becomes unpredictable.

Thus, action being an extensive property, first vanishes causing the system to get highly organized and then causes it to shrink into an infinitesimal point and then re-appears at its maximum magnitude causing the system to become highly unpredictable. Re-organization again starts now in the disintegrated system, causing it to develop its levels of complexity. This unpredictable system again tries to achieve the state of least action and the cycle continues forever. However, this time the course of its development may be entirely different [14, 15, 16, 17, 18, 19].

**Conclusion:** As it was mentioned earlier that this paper does not introduce any new concept or theory, but it tends to look into some of the finer details and their far-reaching consequences in understanding nature in the way it works by not including any kind of pre-assumptions. Nature, in its crude form is very difficult to understand but we must not get carried away by the simplicity of the laws that are presumed to govern nature. As once pointed out by Feynman [20], there is a pleasure in recognizing old things from a new point of view.

The following propositions were made:

1. Action is an extensive property and organization is an intensive property of a dynamical system.
2. Action of a system always assumes a real positive value.
3. At the point of null action the system reaches a dead state and comes in equilibrium with the surrounding.
4. Point of zero action (dead state), point of maximum organization (super-state) and point of maximum action (state of disintegration) are all the same points thus, making the system highly un-predictable (chaotic) at that point. Thus, complexity occurs at the edge of chaos.
5. There always exists a state of competition and conflict between the elements of the system to reach the state of least action since the elements themselves act as constraints for each other.
6. Inspiration has been drawn from game theory to understand the complex nature of the system. Non-cooperative games form the basis of the analysis, since an element within the system calculates its least action irrespective of the other element's strategy. In case internal constraints are present within a system, the geodesic of different groups of elements of the system vary and follow different paths. This produces an illusion of a co-operative game, where a certain number of elements follow the same path, ultimately enhancing the competition between the elements within the system to reach the state of least action.
7. At Nash-equilibrium, the system attains a state of maximum organization (super-state).
8. Organization, disintegration and again, reorganization are instantaneous cyclic processes. At this state, the system behaves chaotically and its future course of evolution is highly sensitive to initial conditions.

These set of propositions are needed to be analysed for highly diverse and complex systems, particularly the open systems, as nature essentially deals with them. New mathematical tools and elegant approaches are required to supplement logical reasoning in order to comprehend the real beauty of nature and natural systems.

**References:**

1. Ville, R., Kaila, I. and Annila, A., "Natural selection for least action" *Proceedings of The Royal Society* **464**, 3055-3070 (2008).
2. Georgiev, G. and Georgiev, I., "The Least Action and the Metric of an Organized system" *open systems and information dynamics* **9,** (4) 371-380 (2002).
3. Neumann, J. von & Morgenstern, O., *Theory of Games and Economic Behavior*, (Princeton University Press, 1944).